\documentclass[useAMS,usenatbib,psfig]{mn2e}

\usepackage{amssymb}
\usepackage{times}
\usepackage{color}
\usepackage{ulem}
\usepackage{graphicx}

\title[A failed-outflow model for TDE]{A failed-outflow model for the UV/optical blackbody emission in tidal disruption events}
\author[X. Cao \& T. Wang]
{Xinwu Cao$^{1,2,3}$ and Tinggui Wang$^{4,5}$\\
$^1$ Shanghai Astronomical Observatory, Chinese Academy of Sciences,
80 Nandan Road, Shanghai, 200030, China; E-mail: cxw@shao.ac.cn\\
$^2$ Key Laboratory of Radio Astronomy, Chinese Academy of Sciences,
210008 Nanjing, China\\
$^3$ University of Chinese Academy of Sciences, Beijing 100049,
China\\
$^4$ CAS Key Laboratory for Research in Galaxies and Cosmology, University of Science and Technology of China, Hefei, Anhui, 230026, China;\\ E-mail: twang@ustc.edu.cn\\
$^5$ School of Astronomy and Space Science, University of Science and Technology of China, Hefei, Anhui, 230026, China}

\date{submitted to MNRAS Letters on \today}
\pagerange{\pageref{firstpage}--\pageref{lastpage}} \pubyear{}

\begin{document}

\maketitle \label{firstpage}

\begin{abstract}
The temperature remains nearly constant while blackbody emission in UV/optical bands declines more than one order of magnitude in some tidal disruption events (TDEs). The physics behind it is still a mystery. A strong outflow can be driven by the radiation of the disc with super-Eddington luminosity. The disc emission drops rapidly to sub-Eddington luminosity, and the gas may fall back to the black hole. An optically thick shell is formed with gas temperature $\la 5\times 10^4$K due to line absorption, which is irradiated by the disc, and is re-emitting UV/optical photons. As the shell moves inwards, the gas at the inner surface of the shell is completely ionized at a certain temperature $\sim 10^4-5\times 10^4$K, which makes the gas optically thin for line absorption, and it therefore falls from the shell. This line absorbing layer acts as temperature regulator, i.e., the gas in the inner shell surface is removed to reduce the shell mass while the temperature is rising, and the decrease of the shell mass (gravity) decelerates the shell till a new balance is achieved between the radiation force and the gravity. This failed-outflow model can naturally explain the declining UV/optical blackbody radiation with constant temperature.
\end{abstract}


\begin{keywords}
accretion, accretion discs---black hole physics---galaxies: nuclei
\end{keywords}

\section{Introduction}\label{intro}

A star passing around a massive black hole within the tidal disruption radius of the hole will be torn apart, which is namely tidal disruption events (TDEs). A fraction of disrupted star mass is accreted onto the black hole through the disc, and evidence of outflows has been observed in some TDEs \citep*[e.g.,][]{2016ApJ...819L..25A,2016ApJ...832L..10R,2017ApJ...837..153A,2018MNRAS.474.3593K}. The mass accretion rate of the disc can be very high initially, and then it declines rapidly in the timescale of $\sim1$~yr, which is a good laboratory for studying the physics of accretion disc. The simple "fallback`` model assuming equal mass in equal energy intervals predicted the mass accretion rate follows $t^{-5/3}$ in TDEs \citep*[][]{1988Natur.333..523R,1989ApJ...346L..13E,1989Natur.340..595P}. However, analysis of the X-ray light curves for a large sample of TDEs showed that most of them can be fitted with power-law, but power-law indices are typically shallower than the canonical $t^{-5/3}$ decay law \citep*[][]{2017ApJ...838..149A}. It was suggested that evolution of a viscous accretion disc may play an important role in late-time TDE light curves \citep*[][]{1990ApJ...351...38C,2009ApJ...700.1047C,2014ApJ...784...87S,2011MNRAS.410..359L,2011ApJ...742...32C,2015ApJ...809..166G,2018MNRAS.tmp.2356B}.

Many TDEs were discovered by the UV/optical and X-ray surveys \citep*[e.g,][]{1999A&A...349L..45K,1999A&A...350L..31G,2008ApJ...676..944G,2009ApJ...698.1367G,2011Sci...333..203B,2011Natur.476..421B,2012MNRAS.420.2684C,2014ApJ...781...59D}. Some of them have been extensively observed in time domain, and the observed UV/optical emission is found to be well fitted by blackbody with temperature $\sim 3-5\times 10^4$~K in several TDEs \citep*[e.g.,][]{2014MNRAS.445.3263H,2016MNRAS.455.2918H,2016MNRAS.462.3993B,2017MNRAS.466.4904B,2018arXiv180802890H}. The temperature remains nearly constant while the luminosity declines more than one order of magnitude \citep*[][]{2016MNRAS.455.2918H,2018arXiv180802890H}. The physics behind it is still a mystery. In ASASSN-14li, the blackbody temperature is $\sim 35000$~K, while the luminosity declines by a factor of 16 over $\sim6$ months \citep*[][]{2016MNRAS.455.2918H}. The X-ray observations with \textit{XMM-Newton} show that the spectra of this source can be fitted with the DISKBB model fairly well, which provides strong evidence of accretion disc origin for X-ray emission \citep*[][]{2018MNRAS.474.3593K}. It implies that the UV/optical blackbody emission should not originate from the accretion disc. The X-ray luminosity at early times is $\sim10^{45}~{\rm erg~s}^{-1}$, which is around the Eddington luminosity of a $10^{6.5}M_\odot$ black hole.

In this work, we suggest a failed-outflow model for the UV/optical blackbody emission in ASASSN-14li. We note that this model is also applicable for variable UV/optical blackbody emission with constant temperature observed in other TDEs.

\section{Model}\label{model}

The model fit to the X-ray spectra of ASASSN-14li indicates the disc temperature is $\sim 50$eV, and the gas of the disc is almost completely ionized at this high temperature \citep*[][]{2018MNRAS.474.3593K}. {For tidal disruption of a solar-type star by a black hole of mass less than $10^7 M_{\odot}$,  the initial accretion rate of the disc formed in the TDE is super-Eddington.} A fraction of the gas may be driven away from the disc surface by the disc radiation to form an outflow. The gas in the outflow is accelerated by the radiation force due to electron scattering of the photons from the disc with super-Eddington luminosity, while outflows may still be driven by radiation force due to line absorption even if the disc emission is sub-Eddington luminosity \citep*[e.g.,][]{2014MNRAS.438.3024L}. When the gas moves outwards to a large distance from the black hole, the temperature of the gas drops to $\la 5\times 10^4$K due to expanding or/and cooling of the gas in the outflow \citep*[e.g.,][]{2010ApJ...724..855C}. The properties of such radiation driven outflows have been studied by numerical simulations \citep*[e.g.,][]{1998MNRAS.295..595P,1999MNRAS.310..476P,2000ApJ...543..686P,2016ApJ...830..125J,2018ApJ...859L..20D}, which show that the outflows usually have a hollow geometry, i.e., the gas in the outflow is expanding outwards with a narrow funnel in the polar direction. The radiation from the inner region of the disc can be well observed without obscured or only slightly obscured by the tenuous gas in the sight line if it is viewed nearly pole-on \citep*[see][for the details]{2018ApJ...859L..20D}.

\begin{figure}
  \includegraphics[width=0.6\textwidth]{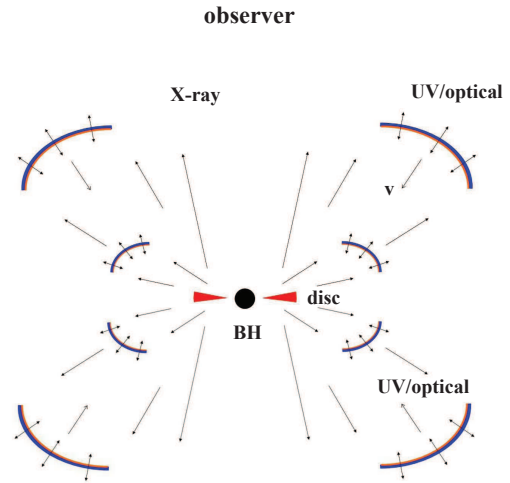}
\caption{Illustration of the failed-outflow model for the UV/optical emission in the TDE. The gas in the failed-outflow falls back to the accretion disc, which forms a shell roughly in equilibrium between the gravity and the radiation force. A thin layer due to line absorption with temperature $\la 5\times10^4$~K is in the inner surface of the shell (orange), which is covered with a thin layer of gas with slightly lower temperature (blue). The shell is optically thick for line/free-free/bound-free absorption. For ASASSN-14li, the melt-down temperature is $\sim 35000$~K as indicated by the blackbody emission observed in UV/optical wavebands. The shell moves slowly towards the black hole with decreasing disc radiation. \label{illustr} }
\end{figure}

In the outer region of the outflow, the line opacity becomes dominant over the electron scattering opacity. The line opacity can be several orders of magnitude higher than that of the electron scattering for the gas with temperature $\sim 10^4-5\times10^4$~K, which drops sharply at temperature $\la 10^4$~K \citep*[][]{1982ApJ...259..282A}. An optically thick shell due to line absorption is formed in the outflow when the gas temperature drops to a certain temperature $\la 5\times10^4$~K (it could be lower if the photo-ionization by the X-ray radiation is properly considered).  If the surface mass density of the shell is sufficiently high and it is therefore optically thick for free-free/bound-free processes, the incident photons from the disc are thermalized and therefore the outgoing flux from the shell is nearly blackbody emission,


The X-ray spectra of ASASSN-14li imply that the source may probably be viewed nearly pole-on \citep*[][]{2018MNRAS.474.3593K}. For a TDE, the mass accretion rate declines rapidly, and it will soon be sub-Eddington. This leads to a lower radiation force, which makes the gas unable to overcome the gravitational barrier.
The shell will stop going outwards and then fall back to the accretion disc, namely a failed-outflow.  An illustration of the fail-outflow model for TDEs is plotted in Figure \ref{illustr}.

In the inner surface of the shell, the gas is ionized by the X-ray photons from the accretion disc. The opacity of the gas is dominated by the electron scattering, and the gas in this region falls inwards rapidly in the case of the disc with sub-Eddington luminosity. In the layer behind this photo-ionized gas, the photo-ionization parameter 
decreases to a low value, and therefore line absorption becomes important, of which the opacity can be hundreds (or even higher) times the electron scattering cross-section \citep*[][]{2000ApJ...543..686P}. It was found that the line opacity remains constant (roughly), when the gas temperature in the range of $\sim 10^4-5\times 10^4$K, and then drops sharply when the temperature is beyond this range  \citep*[][]{1982ApJ...259..282A}. The gas at the inner surface of the shell is completely ionized and its line opacity decreases sharply at a certain temperature lower than $\sim5\times10^4$~K, which is roughly the temperature of the inner surface of the shell (hereafter it is referred to as ``melt-down temperature"). In principle, the melt-down temperature can be calculated with proper consideration of detailed atomic processes, which is beyond the scope of this work. The incident photons in the range of the frequencies of absorption lines are absorbed in a thin layer at the shell surface, which is sensitive to the radial velocity gradient \citep*[e.g.,][]{1982ApJ...259..282A}. Most continuum emission beyond these ranges of the absorption line frequencies are absorbed in the gas of the shell by free-free/bound-free processes. If the optical depth of the shell is much greater than unity, the shell emits blackbody radiation with melt-down temperature.
 The radiation force exerted on the shell is roughly in equilibrium with the gravity for a slowly shrinking shell. As the disc radiation decreases, the shell moves towards the black hole. The shell temperature remains at the melt-down temperature so that the shell surface is still optically thick for line absorption. This temperature is well determined by the observed blackbody radiation in the UV/optical wavebands.


In our model, we assume that the optically thick shells formed in the outflows are irradiated by the accretion disc, so the blackbody emission in UV/optical wavebands is radiated from the two surfaces of the shells. For an optically thick shell, the UV/optical flux from unit area of the outer surface of the shell is
\begin{equation}
f_{\rm rad,out}=\sigma T_{\rm sh,out}^4\simeq {\frac {4\sigma}{3\tau}}T_{\rm sh,in}^4,\label{f_rad_out}
\end{equation}
where $\tau=\kappa\Sigma$ is the optical depth of the shell. The flux emitted from the inner surface is
\begin{equation}
f_{\rm rad,in}=\sigma T_{\rm sh,in}^4.\label{f_rad_in}
\end{equation}
The energy conservation for the irradiated shell requires
\begin{equation}
f_{\rm X}\simeq f_{\rm rad,in}+f_{\rm rad,out},
\label{f_x}
\end{equation}
where $f_{\rm X}$ is the incident X-ray flux. Combining Equations (\ref{f_rad_out}) and (\ref{f_rad_in}), we have
\begin{equation}
f_{\rm rad,in}\simeq {\frac {3\tau}{4}}f_{\rm rad,out},
\label{f_in_out}
\end{equation}
which indicates
\begin{equation}
f_{\rm rad,in}\gg f_{\rm rad,out}~~~ {\rm if}~~~ \tau\gg 1.
\label{f_in_out2}
\end{equation}
This implies that most observed UV/optical emission originates from the inner surface of the shell provided the source is viewed nearly pole-on and the covering factor $f_{\rm cov}$ is small (see the illustration in Figure 1, the UV/optical photons emitted from the outer surface of the shell above the disc and the inner surface of the shell below the disk are directly detected by a distant observer). We note that the size of the accretion disk is very small, usually smaller than one hundred gravitational radii for a solar-like star disrupted by a $10^{6.5}M_\odot$ black hole \citep*[][]{1988Natur.333..523R}. It is much smaller than $R_{\rm sh}$, which is to be verified later, so the obscuration by the disk may be negligible at the early times of observation when the shell is still far away from the black hole. It is found that $T_{\rm sh,in}\simeq(3\tau/4)^{1/4}T_{\rm sh,out}$ from Equation (\ref{f_rad_out}). The temperature $T_{\rm sh,out}$ of the outer surface of the shells is lower than the inner surface temperature $T_{\rm sh,in}$, but not much lower if the optical depth $\tau$ is up to ten.


The UV/optical luminosity of the shells is
\begin{equation}
L_{\rm UV}\sim f_{\rm cov}L_{\rm disc},
\label{uv_optical_lum}
\end{equation}
where $f_{\rm cov}$ is the covering factor of the shells, and $L_{\rm disc}$ is the disc luminosity, which is dominant in X-ray bands, and hereafter $L_{\rm disc}\simeq L_{\rm X}$ is assumed. As the shell is irradiated by the accretion disc and to re-radiate in UV/optical bands, we estimate the covering factor as $f_{\rm cov}\sim L_{\rm UV}/L_{\rm X}$. The X-ray luminosity $L_{\rm X}\sim 10^{45}$~erg~s$^{-1}$ at the early time \citep*[][]{2018MNRAS.474.3593K}, and one can derive $f_{\rm cov}\la 0.1$ initially.
The UV/optical luminosity is
\begin{displaymath}
L_{\rm UV}=4\pi R_{\rm sh}^2f_{\rm cov}(f_{\rm rad,in}+f_{\rm rad,out})
\end{displaymath}
\begin{equation}
~~~~~~~~\simeq 4\pi R_{\rm sh}^2f_{\rm cov}f_{\rm rad,in}=4\pi R_{\rm sh}^2f_{\rm cov}\sigma T_{\rm sh,in}^4, \label{uv_optical_lum3}
\end{equation}
where $R_{\rm sh}$ is the distance of the shell from the black hole, and the approximation $f_{\rm rad,in}\gg f_{\rm rad,out}$ is adopted for the optically thick shell (see Equations \ref{f_in_out} and \ref{f_in_out2}).

It is found that the observed light curve in the UV/optical wavebands can be well described by
\begin{equation}
L_{\rm UV}=6.84\times10^{43} \exp\left(-{\frac {t}{60}}\right)~~~{\rm erg~s}^{-1},
\label{uv_optical_lum4}
\end{equation}
where $t$ is in unit of day \citep*[][]{2016MNRAS.455.2918H}. The UV/optical emission of ASASSN-14li is well fitted by a blackbody with roughly constant temperature $35000$~K, i.e., $T_{\rm sh,in}=35000$~K, in the period of $\sim200$ days, during which the UV/optical emission declines to $\sim 1/16$ of its initial value. Combining Equations (\ref{uv_optical_lum3}) and (\ref{uv_optical_lum4}), we have
\begin{displaymath}
r_{\rm sh}={\frac {R_{\rm sh}}{R_{\rm g}}}
\end{displaymath}
\begin{equation}
=2.1\times 10^{4}f_{\rm cov}^{-1/2}
\left({\frac {M_{\rm bh}}{10^6M_\odot}}\right)^{-1}
\left({\frac {T_{\rm sh,in}}{10^4}}\right)^{-2}\exp\left(-{\frac {t}{120}}\right),
\label{r_sh}
\end{equation}
where
\begin{equation}
r_{\rm sh}={\frac {R_{\rm sh}}{R_{\rm g}}}, ~~~{\rm and}~~~R_{\rm g}={\frac {GM_{\rm bh}}{c^2}}.
\end{equation}
For ASASSN-14li, its black hole mass $M_{\rm bh}=10^{6.5}M_\odot$ \citep*[][]{2016Sci...351...62V}, which is very close to other estimates \citep*[see][and the references therein]{2014MNRAS.445.3263H,2016MNRAS.455.2918H}. The shell is located at $R_{\rm sh}\sim 1700~R_{\rm g}$ at the early time (i.e., $t=0$), if $f_{\rm cov}=0.1$ is adopted. At this distance, the free-fall velocity, $V_R^{\rm ff}=(GM_{\rm bh}/R_{\rm sh})^{1/2}=cr_{\rm sh}^{-1/2}\simeq 7.3\times10^8$~cm~s$^{-1}$. We can estimate the velocity of the shell moving towards the black hole,
\begin{displaymath}
V_R=\left |{\frac {dR_{\rm sh}}{dt}}\right |=9.65\times 10^{-8}R_{\rm sh}=4.51\times10^4r_{\rm sh}
\label{v_r}
\end{displaymath}
\begin{equation}
~~~~~=2.44\times 10^7f_{\rm cov}^{-1/2}\exp\left(-{\frac {t}{120}}\right)~{\rm cm~s}^{-1},
\end{equation}
which is always much lower than the free-fall velocity (about one order of magnitude lower than the free-fall velocity at early times, and much lower at late times, because $V_R/V_{R}^{\rm ff}\propto R_{\rm sh}^{3/2}$). It indicates that the equilibrium between the radiation force and the gravity is a good approximation, i.e.,
\begin{equation}
{\frac {GM_{\rm bh}\Sigma}{R_{\rm sh}^2}}\sim{\frac {f_{\rm X}}{c}}-{\frac {f_{\rm rad,out}}{c}}+{\frac {f_{\rm rad,in}}{c}}\sim{\frac {2f_{\rm X}}{c}},
\label{sigma_sh}
\end{equation}
for an optically thick shell, where $\Sigma$ is the surface mass density of the shell, and relations (\ref{f_x}) and (\ref{f_in_out2}) are used. This provides estimates of the surface mass density and total mass of the shells,
\begin{equation}
\Sigma\sim 4f_{\rm cov}^{-1}\left({\frac {L_{\rm UV}}{10^{44}}}\right)\left({\frac {M_{\rm bh}}{10^6M_\odot}}\right)^{-1}~{\rm g~cm^{-2}},
\label{sigma_sh2}
\end{equation}
and
\begin{equation}
{\frac {M_{\rm sh}}{M_\odot}}\sim 0.355f_{\rm cov}^{-1}\left({\frac {T_{\rm sh,in}}{10^4}}\right)^{-4}\left({\frac {L_{\rm UV}}{10^{44}}}\right)^2\left({\frac {M_{\rm bh}}{10^6M_\odot}}\right)^{-1}, \label{m_sh}
\end{equation}
where Equations (\ref{uv_optical_lum}) and (\ref{uv_optical_lum3}) are used. At $t=0$, we find that $M_{\rm sh}\sim3.5\times 10^{-3}~M_\odot$.

Assuming the gas pressure gradient to be in equilibrium with the gravity in the rest frame of the shell, we have
\begin{equation}
{\frac {dp}{dR}}=-{\frac {GM_{\rm bh}\rho}{R^2}},
\label{l_sh}
\end{equation}
which can be approximates as
\begin{equation}
{\frac {dp}{dR}}\sim{\frac {\Delta p}{l_{\rm sh}}}\sim -{\frac {\rho kT_{\rm in}}{\mu m_{\rm p}l_{\rm sh}}}\sim-{\frac {GM_{\rm bh}\rho}{R_{\rm sh}^2}},
\label{l_sh2}
\end{equation}
where $l_{\rm sh}$ the shell thickness. So, the thickness can be estimated as
\begin{equation}
{\frac {l_{\rm sh}}{R_{\rm sh}}}\sim3.94\times 10^{-5}f_{\rm cov}^{-1/2}\left({\frac {T_{\rm sh,in}}{10^4}}\right)^{-1}\left({\frac {L_{\rm UV}}{10^{44}}}\right)^{1/2}\left({\frac {M_{\rm bh}}{10^6M_\odot}}\right)^{-1},
\label{l_sh3}
\end{equation}
or
\begin{equation}
l_{\rm sh}\sim 1.48\times10^{11}f_{\rm cov}^{-1}\left({\frac {T_{\rm sh,in}}{10^4}}\right)^{-3}\left({\frac {L_{\rm UV}}{10^{44}}}\right)\left({\frac {M_{\rm bh}}{10^6M_\odot}}\right)^{-1}~{\rm cm}.
\label{l_sh4}
\end{equation}
Combining Equations (\ref{sigma_sh2}) and (\ref{l_sh4}), we obtain
\begin{equation}
\rho\sim{\frac {\Sigma}{l_{\rm sh}}}=2.7\times10^{-11}\left({\frac {T_{\rm sh,in}}{10^4}}\right)^{3}~{\rm g~cm^{-3}},
\label{rho}
\end{equation}
which implies the opacity of the shell remains constant for ASASSN-14li as its temperature is almost unchanged when the UV/optical luminosity declines more than one order of magnitude \citep*[][]{2016MNRAS.455.2918H}. We can estimate the optical depth of the shell,
\begin{displaymath}
\tau=\kappa\Sigma\sim 3.52f_{\rm cov}^{-1}\left({\frac {L_{\rm UV}}{10^{44}}}\right)\left({\frac {M_{\rm bh}}{10^6M_\odot}}\right)^{-1}
\end{displaymath}
\begin{equation}
~~~~~=3.52\left({\frac {L_{\rm X}}{10^{44}}}\right)\left({\frac {M_{\rm bh}}{10^6M_\odot}}\right)^{-1}, \label{tau_sh}
\end{equation}
where the opacity $\kappa=[\kappa_{\rm R}(\kappa_{\rm es}+\kappa_{\rm R})]^{1/2}$ ($\kappa_{\rm es}$ is the electron scattering opacity, and the Rosseland mean opacity $\kappa_{\rm R}=5\times10^{24}\rho T^{-7/2}~{\rm cm^2~g^{-1}}$), and $T_{\rm sh,in}=35000$~K is adopted for ASASSN-14li. It is found that $\tau\sim8$ when $t=0$. At late times, the X-ray luminosity roughly declines to one-fourth of its initial luminosity in a period of time during which its UV/optical luminosity drops to 1/16 of its initial value \citep*[][]{2018MNRAS.474.3593K}. It implies that the covering factor $f_{\rm cov}$ at late times is about one-fourth of its initial value. With Equation (\ref{tau_sh}), one can estimate that the optical depth $\tau\sim2$ at the late times, which is still optically thick.

The cooling timescale of the shell due to the UV/optical radiation is
\begin{displaymath}
\tau_{\rm cooling}\sim {\frac {3kT_{\rm sh}}{2\mu m_{\rm p}}}{\frac {M_{\rm sh}}{L_{\rm UV}}}
\end{displaymath}
\begin{equation}
~~~~~=14.8f_{\rm cov}^{-1}\left({\frac {T_{\rm sh}}{10^4}}\right)^{-3}\left({\frac {L_{\rm UV}}{10^{44}}}\right)\left({\frac {M_{\rm bh}}{10^6M_\odot}}\right)^{-1}~{\rm second}.
\label{tau_cool}
\end{equation}
It is obvious that the cooling is very efficient, i.e., the UV/optical emission varies with the disc radiation nearly simultaneously.




\section{Discussion}\label{discussion}

The inner surface of the shell is irradiated by the X-ray radiation of the disc. The radiation force is roughly in balance with the gravity of the black hole. If there is a small perturbation making the shell shift a small distance to the black hole, the X-ray flux increases due to a smaller distance to the accretion disc, which leads to an increase of the radiation force against the gravity. This also makes much gas being ionized in the inner surface of the shell, and falls onto the accretion disc, which reduces the mass of the shell. The shell is therefore pushed outwards due to these two effects. If the shell is shifted outwards, the radiation force decreases as a longer distance to the black hole, and the gravity dominates over the radiation force. The shell will move back to the balance point. The self-regulation mechanism described above shows that the structure of the shell with force balance is stable, which allows the shell moves smoothly to the black hole.

We find that the shell mass is required to decrease with decreasing UV/optical luminosity by the force balance between radiation force and the gravity (see Equation \ref{m_sh}), which implies a mechanism needed to reduce the shell mass when the accretion luminosity declines. The gravity becomes larger than the radiation force if the X-ray radiation from the disc decreases, which makes the shell accelerate towards the black hole. The gas at the inner surface of the shell is photo-ionized and the temperature of gas in the inner region of the line-absorption layer may be heated to a temperature higher than the melt-down temperature, which makes the gas completely ionized. The radiation force exerted on the ionized gas is caused by the electron scattering of the disc, which is substantially lower than the line force \citep*[][]{2000ApJ...543..686P}. The force balance of the line force and gravity of the gas at the inner surface is broken with declining disc luminosity, which leads to rapid acceleration of the gas to the black hole at the inner surface of the shell, and the gas is stripped from the shell. Therefore, this line absorbing layer acts as temperature regulator, i.e., the gas in the inner shell surface is removed to reduce the shell mass while the temperature is rising, and the decrease of the shell mass (gravity) decelerates the shell till a new balance is soon achieved between the radiation force and the gravity, as the timescale of atomic processes is always very short. We note that the photo-ionization parameter remains constant ($\xi=L_{\rm X}/n_{\rm e}R_{\rm sh}^2\sim3$) in the inner surface of the shell, because the gas density of the shell does not vary with time (see Equation \ref{rho}), which implies a constant degree of ionization for inner surfaces of shrinking shells. With this ionization parameter, we expect most heavy elements are partially ionized. Thus soft X-rays are absorbed via the photoelectric process in a relatively thin layer with column number density $\sim 10^{22-23}{\rm cm}^{-2}$ on the illuminating face. The temperature of the gas at the inner surface of the shell is well regulated at a certain temperature of $\sim 10^4-5\times10^4$K. This can naturally explain the constant temperature of the blackbody emission observed in UV/optical bands \citep*[e.g.,][]{2018arXiv180802890H}.

The shell mass decreases with the UV/optical luminosity (see Equation \ref{m_sh}), which implies that the gas stripped from the shell may fall onto the disc and finally be accreted by the black hole. As the mass of the shell $M_{\rm sh}$ decreases rapidly with decreasing UV/optical luminosity $L_{\rm UV}$ (see Equation \ref{m_sh}), almost all initial mass of the shell (at $t=0$) will be finally accreted by the black hole. The additional energy released due to falling-back shell is $\sim\eta M_{\rm sh}c^2\sim6.3\times10^{50}$~erg~s$^{-1}$, assuming an efficiency $\eta=0.1$, which is comparable with, $\sim7\times10^{50}$erg, the total energy derived by integrating the X-ray and UV/optically light curves \citep*[][]{2016MNRAS.455.2918H}. Analysis of the X-ray light curves for a large sample of TDEs showed that most of them can be fitted with power-law, but power-law indices are typically shallower than the canonical $t^{-5/3}$ decay law \citep*[][]{2017ApJ...838..149A}. The gas in the failed-outflow may re-fuel the accretion disc, which may help alleviate this discrepancy. The detailed model calculations of the light curve of the disc with failed-outflows are beyond the scope of this work.

It is found that the X-ray luminosity declines to about one-fourth of it at early times in a period of $\sim200$~days, while it is about 1/16 for  the UV/optical emission in the same period of time \citep*[][]{2016MNRAS.455.2918H,2018MNRAS.474.3593K}. This implies that the covering factor $f_{\rm cov}$ decreases with decreasing disc radiation (see Equation \ref{uv_optical_lum}), or/and the disc radiation is anisotropic ($\propto \cos\theta$ for a parallel plane) . The gas in the outflow may be slowly rotating, because the gas was initially driven from a rotating accretion disc, which implies that the angular velocity of the gas may be higher when it falls closer to the black hole. The centrifugal force may decelerate the gas in radial direction, while vertical component of the gas velocity is not affected. This would make the gas move more close to the equatorial plane, which leads to a smaller covering factor. The detailed calculations of this issue is desired most probably by numerical simulations. The present model calculations are carried out mostly for the UV/optical blackbody emission in ASASSN-14li. We note that this model is also applicable for the UV/optical blackbody emission observed in other TDEs \citep*[][]{2018arXiv180802890H}.

Our model calculations show that the optical depth of the shell declines with time (see discussion in Section \ref{model}), which seems to suggest that the shell will soon be optically thin when it approaches more closer to the black holes. This can be tested by further UV/optical observations. An alternative possibility is that the obscuration of the disc may be important when the shell shrinks to a small distance comparable with the radius of the disc. Considering a viscous accretion disc may expand with time \citep*[][]{2002apa..book.....F}, we find that, the covering factor of the shell to the X-ray radiation may still be large (at least larger than the estimate from the ratio of $L_{\rm UV}/L_{\rm X}$), and the decline of the UV/optical emission may be partly caused by the obscuration of the UV/optical emission from the inner surface of the shell by the expanding accretion disc, i.e., the intrinsic UV/optical emission may not drop so much as observed.

\section*{Acknowledgments}

 This work is supported by the NSFC (grants 11773050 and 11833007), and the CAS grant (QYZDJ-SSW-SYS023).

\end{document}